\documentstyle[12pt,twoside]{article}

\def\greaterthansquiggle{\raise.3ex\hbox{$>$\kern-.75em\lower1ex\hbox{$\sim$}}}
\def\lessthansquiggle{\raise.3ex\hbox{$<$\kern-.75em\lower1ex\hbox{$\sim$}}}
\newcommand{\beq}{\begin{equation}}
\newcommand{\eeq}{\end{equation}}
\newcommand{\beqa}{\begin{eqnarray}}
\newcommand{\eeqa}{\end{eqnarray}}
\newcommand{\beqan}{\begin{eqnarray*}}
\newcommand{\eeqan}{\end{eqnarray*}}
\newcommand{\ba}{\begin{array}}
\newcommand{\ea}{\end{array}}

\def\nz{\ifmmode {I\hskip -3pt N} \else {\hbox {$I\hskip -3pt N$}}\fi}
\def\zz{\ifmmode {Z\hskip -4.8pt Z} \else
       {\hbox {$Z\hskip -4.8pt Z$}}\fi}
\def\qz{\ifmmode {Q\hskip -5.0pt\vrule height6.0pt depth 0pt
       \hskip 6pt} \else {\hbox
       {$Q\hskip -5.0pt\vrule height6.0pt depth 0pt\hskip 6pt$}}\fi}
\def\rz{\ifmmode {I\hskip -3pt R} \else {\hbox {$I\hskip -3pt R$}}\fi}
\def\cz{\ifmmode {C\hskip -4.8pt\vrule height5.8pt\hskip 6.3pt} \else
       {\hbox {$C\hskip -4.8pt\vrule height5.8pt\hskip 6.3pt$}}\fi}

\newtheorem{theorem}{Theorem}

\newtheorem{lemma}{Lemma}

\normalsize
\begin{document}

\begin{titlepage}
\begin{center}
{\Large \bf 
Quantum Dynamical Entropy of Spin Systems
}\\[24pt]
Takayuki Miyadera\\
National Institute of Advanced Industrial Science and 
Technology (AIST),\\
Chiyoda-ku, Tokyo 101-0021, Japan\\
e-mail: miyadera-takayuki@aist.go.jp \\
and \\
Masanori Ohya\\
Department of Information Sciences \\
Tokyo University of Science \\
Noda City, Chiba 278-8510,
Japan
\\
e-mail: ohya@is.noda.tus.ac.jp
\vfill
{\bf Abstract} \\
\end{center}
We investigate a quantum dynamical entropy of one-dimesional quantum spin 
systems. We show that 
the dynamical entropy is bounded from above by a quantity  
which is related with group velocity determined by 
the interaction and mean entropy of the state.
\vfill
\end{titlepage}

\section{Introduction}
The classical dynamical entropy, formulated by Kolmogorov and Sinai, 
is a powerful and sophisticated tool to classify dynamical systems
by characterizing their chaotic property. It can be regarded as a 
quantity which represents an optimal rate of entropy production
by successive observations of a dynamical system. As for quantum dynamical 
systems, there have been several attempts to define quantum 
versions of dynamical entropy \cite{CS,CNT,AF,AFbook,MO,AOW,KOW}. 
The differences and 
similarities among them were discussed in \cite{AOW2,AN}.
In spite of a lot of efforts, because of their difficulties to calculate,
only a few models, like solvable models in quantum optics \cite{KOW},
non-commutative shift \cite{AN}, 
noncommutative Cat map \cite{AFTA,FT}
or quasi free Femionic systems \cite{AFbook}
 have been    
discussed in each formulation, and to the 
authors' knowledge, few nontrivial physical 
models have been ever treated. 
In the present paper, we discuss the quantum dynamical 
entropy defined by Alicki and Fannes \cite{AF,AFbook}
for one-dimensional quantum spin systems. 
We give an upper bound for the quantum dynamical 
entropy of quantum spin systems, 
which is related with mean entropy and
group velocity. 
The upper bound does not depend upon the details of 
the Hamiltonian, and is considered to be rather general one.
The paper is organized as follows.
In section \ref{sect:qde}, we briefly review the 
quantum dynamical entropy with stress on 
its physical interpretation. In section \ref{sect:1d},
the one-dimensional quantum spin system for which we estimate 
the dynamical 
entropy is introduced. In section \ref{sect:sub}, 
we introduce a norm dense $*$-subalgebra on the 
spin system to define the quantum dynamical entropy.
In section \ref{sect:bound},
the main theorem, an upper bound for the quantum dynamical
entropy of the spin systems is obtained.
\section{Quantum Dynamical Entropy}\label{sect:qde}
In this section we briefly review what the quantum dynamical entropy 
defined by Alicki and Fannes \cite{AF,AFbook}
is. 
Suppose there exists a $\mbox{C}^*$ dynamical system 
$({\cal A},\alpha,\omega)$ where ${\cal A}$ is a $\mbox{C}^*$ algebra,
$\alpha$ is a $*$-automorphism on ${\cal A}$ and $\omega$ is an
$\alpha$-invariant state. Let us remind that the classical 
dynamical entropy can be regarded as an entropy production 
rate by successive observations procedure. The kind of 
the observations is to measure in what partition a trajectory 
locates. The philosophy of defining a quantum version of dynamical 
entropy is similar. We consider successive quantum measurements
on the system and discuss how fast the entropy can grow.
In quantum case, a measurement (observation) process is 
nothing but an interaction between the system and 
an apparatus. The process is described mathematically by
a transition expectation \cite{OP}:
\begin{eqnarray}
E:{\cal B} \otimes {\cal A} \to {\cal A}.
\end{eqnarray}
Here ${\cal B}$ represents an observable algebra of 
an apparatus. Since it is not natural to assume 
that we can perform any kinds of measurements, the forms of the 
above transition expectation should be restricted.
First, since we can not count infinity, the dimension of 
the observable algebra of the apparatus 
should be finite. Thus 
${\cal B}$ must be $d$-dimensional matrix algebra $(d=1,2,\cdots)$.
Moreover, the interaction 
process cannot be too sharp.
(If we allow too sharp observations, we always obtain 
infinite dynamical entropy for type III and type $\mbox{I}_{\infty}$ 
von Neumann algebra \cite{FT}.) This restriction is 
realized by introducing a dense subalgebra ${\cal A}_0$ of ${\cal A}$.
That is, the restriction of  $E$ 
on ${\cal B} \otimes {\cal A}_0$ 
must have its image in ${\cal A}_0$,
\begin{eqnarray}
E:{\cal B} \otimes {\cal A}_0 \to {\cal A}_0.
\end{eqnarray}
The successive measurements procedure can be 
described by the following quantum Markov chain \cite{AFL},
\begin{eqnarray}
\alpha\circ E \circ (\mbox{id} \otimes \alpha\circ E)
\circ (\mbox{id}\otimes \mbox{id} \otimes \alpha\circ E)
\circ \cdots (\mbox{id}\otimes \cdots \mbox{id} \otimes
\alpha\circ E):
{\cal B}^{\otimes^n} \otimes {\cal A} \to 
{\cal A}.
\end{eqnarray}
To assure this procedure is also 
the type
${\cal B}^{\otimes^n}\otimes {\cal A}_0 \to {\cal A}_0$,
${\cal A}_0$ should be chosen as an $\alpha$-invariant subalgebra.
In addition, Alicki and Fannes impose externality on the map $E$, that is, 
the map admits no nontrivial decompositions in sums of 
completely positive maps. 
Thanks to Kraus representation theorem, such an extreme map 
has a representation by the 
corresponding operational partition of unity
$\{ x_i\}_{i=1}^Z \subset {\cal A}_0$ such that 
$\sum_{i}x_i^* x_i ={\bf 1}$, as
\begin{eqnarray}
E(A)= \sum_{i,j} x^*_i A_{ij} x_j,
\end{eqnarray}
where $A_{ij}$ is an $i,j$ component of 
$M_Z({\cal A}_0) =M_Z({\bf C})\otimes {\cal A}_0$.
On this setting, a finitely correlated state \cite{FNW} $\rho_n$ 
over ${\cal B}^{\otimes^n}$ is 
given by 
\begin{eqnarray}
&&\rho_n(B_1\otimes B_2 \otimes \cdots B_n)
:= 
\nonumber \\
&&\omega\circ\alpha\circ E \circ (\mbox{id} \otimes \alpha\circ E)
\circ \cdots (\mbox{id}\otimes \cdots \mbox{id} \otimes
\alpha\circ E)
(B_1 \otimes B_2 \otimes \cdots B_n \otimes {\bf 1}).
\end{eqnarray}
The dynamical entropy is defined by 
the supremum of mean von Neumann entropy $S(\rho_n)$ of $\rho_n$ 
for all the possible transition expectations 
within the restriction.
For $M=1,2,\cdots,$,
let us define $Z^M \times Z^M$ matrix $\rho(\{x_i\}_i,M)$ by 
its components $\omega(x^*_{i_0}\alpha(x^*_{i_1})
\cdots \alpha^{M-1}(x^*_{i_{M-1}})
\alpha^{M-1}(x_{j_{M-1}})\cdots\alpha(x_{j_1})x_{j_0})$
for $i_s,j_s =1,\cdots,Z$ and $s=1,\cdots,M$.
By use of this expression, one can write the 
dynamical entropy as 
\begin{eqnarray}
 h(\omega,\alpha,{\cal A}_0):=\sup \left\{ 
\limsup_M 
\frac{1}{M}S(\rho(\{x_i\}_i,M))|\ \{x_i\}_{i=1}^Z \subset {\cal A}_{0}; 
Z=1,
2,\cdots;
\sum_{i=1}^Z x_i^*x_i ={\bf 1}\right\}.
\end{eqnarray}
\section{One-dimensional Spin Systems}\label{sect:1d}
We deal with one-dimensional two-way infinite quantum
spin systems. 
To each site $x\in {\bf Z}$ a Hilbert space ${\cal H}_{x}$ which is
isomorphic to ${\bf C}^{N+1}$ is attached and the observable algebra at site 
$x$ is a matrix algebra on ${\cal H}_{x}$ 
which is denoted by ${\cal A}(\{x\})$. 
The observable algebra on a finite set $\Lambda \subset {\bf Z}$ is a matrix
algebra 
on $\otimes _{x\in \Lambda }{\cal H}_{x}$ and denoted by ${\cal A}(\Lambda )$%
. Natural identification can be used to derive an inclusion property ${\cal %
A}(\Lambda _{1})\subset {\cal A}(\Lambda _{2})$ for $\Lambda _{1}\subset
\Lambda _{2}$. The total observable algebra is a norm completion of sum
of the finite region observable algebra, ${\cal A}:=\overline{\cup _{\Lambda
:\mbox{\small{finite}}}{\cal A}(\Lambda )}^{\Vert \Vert }$, which becomes a $%
\mbox{C}^{\ast }$ algebra. We write the translation automorphism 
as $\tau_x\ (x\in {\bf Z})$. (For detail, see \cite{BR}.) 

To discuss the dynamics, we need, $\{\alpha_t\}_{t \in {\bf R}}$,
 a one-parameter $\ast $%
-automorphism group on ${\cal A}$, which we assume is induced by a local
interaction. That is, we assume that for each finite subset $\Lambda
\subset {\bf Z}$, a potential $\Phi(\Lambda)\in {\cal A}(\Lambda)$ is 
defined. The translational invariance $\tau_x(\Phi(\Lambda))
=\Phi(\Lambda+x)$ for all $x \in {\bf Z}$ and finite $\Lambda \in {\bf Z}$ 
is assumed.
The potential satisfies the locality condition, that is, 
there exists $\lambda >0$ satisfying the following inequality,
\begin{eqnarray}
\Vert \Phi \Vert_{\lambda}:=
\sup\left\{
|X|(N+1)^{2|X|}e^{\lambda D(X)}\Vert \Phi(X)\Vert
|\ X \mbox{ is a finite subset of }{\bf Z}
\right\}<\infty,
\label{local}
\end{eqnarray}
where $D(X):=\sup_{x,y \in X} |x-y|$.
(This condition is used later to define a group velocity.)
Local Hamiltonian with respect to 
a finite region $\Lambda \in {\bf Z}$ 
is defined as 
\begin{eqnarray}
H_{\Lambda}:=\sum_{V\subset \Lambda} \Phi(V),
\end{eqnarray}
which induces a one-parameter $*$-automorphism 
group $\alpha^{\Lambda}_t\ 
(t \in {\bf R})$ 
by 
\begin{eqnarray}
\alpha^{\Lambda}_t(A):=e^{iH_{\Lambda}t}Ae^{-iH_{\Lambda}t}
\end{eqnarray}
for each local element $A\in {\cal A}$. 
Thanks to the locality of the interaction (\ref{local}),
the infinite volume limit for $\Lambda$ converges to a
$*$-automorphism $\alpha_t$ in norm topology, i.e., 
\begin{eqnarray}
\lim_{\Lambda \to {\bf Z}}
\Vert \alpha_t(A)-\alpha^{\Lambda}_t(A)\Vert =0
\end{eqnarray}
holds \cite{BR}.
\section{Dynamical Entropy of Quantum Spin Systems}\label{sect:sub}
Let us consider the dynamical entropy of the spin systems.
We fix a time-invariant state $\omega$ and express simply $\alpha:=
\alpha_{t=1}$. We investigate the dynamical entropy for discrete
dynamical system $\left( {\cal A},\alpha,\omega\right)$.
To calculate dynamical entropy, we must choose a 
natural time-invariant subalgebra ${\cal A}_0 \subset {\cal A}$ for 
partitions of unity. Although the most natural
choice in our spin system case seems to be a norm dense subalgebra which 
is composed of strictly local objects, 
\begin{eqnarray}
{\cal A}_{loc}:=\cup_{\Lambda \subset
{\bf Z}: \mbox{finite}} {\cal A}(\Lambda).
\end{eqnarray}
This subalgebra, however, is not time invariant and 
therefore cannot be a candidate. 
Let us define a slightly larger algebra composed of
exponentially localized objects.
For a quasilocal object $A\in {\cal A}$,
how strong it lives on a site $x$ can be measured by 
a quantity,
\begin{eqnarray}
F_x(A):=\sup_{a\in {\cal A}(\{0 \}),a\neq 0}\left( 
\frac{\Vert [A,\tau_x(a)]\Vert}{\Vert a\Vert}
\right).
\end{eqnarray}
The set of exponentially localized objects is defined as
\begin{eqnarray}
{\cal A}_{\exp}:=
\{ A \in {\cal A}|\ ^{\exists}\mu >0\ 
\mbox{such that}\  
\lim_{|x|\to \infty}e^{\mu |x|} F_x(A)=0  \}.
\end{eqnarray}
This set becomes a $*$-subalgebra. In fact, one can easily check that 
for any $c_1, c_2 \in {\bf C}$ and $A_1,A_2 \in {\cal A}_{\exp}$,
\begin{eqnarray}
F_x(c_1A_1+c_2A_2)\leq |c_1| F_x(A_1)+|c_2|F_x(A_2)
\end{eqnarray}
and hence ${\cal A}_{\exp}$
 is closed with linear combination.
In addition, for $A, B\in {\cal A}_{\exp}$, the equation,
\begin{eqnarray}
F_x(AB)&=&\sup_{a\in {\cal A}(x),a\neq 0}\left( 
\frac{\Vert [AB,a]\Vert}{\Vert a\Vert}
\right)\nonumber \\
&=&
\sup_{a\in {\cal A}(x),a\neq 0}\left( 
\frac{\Vert [A,a]B +A[B,a]\Vert}{\Vert a\Vert}
\right) 
\nonumber \\
&\leq&
F_x(A)\Vert B\Vert +\Vert A\Vert F_x(B)
\end{eqnarray}
leads that $AB \in {\cal A}_{\exp}$.
Finally, obviously
$F_x(A)=F_x(A^*)$ holds.
Moreover one can show the following theorem:
\begin{theorem}
The set of exponentially localized objects ${\cal A}_{\exp}$
is an $\alpha_t$-invariant $*$-subalgebra.
\end{theorem}
{\bf Proof:}
\\
According to Proposition 6.2.9. in \cite{BR}, for any $\Lambda \subset 
{\bf Z}$, 
for all $a_y\in {\cal A}(\{y\})$ and $B \in {\cal A}_{\exp}$,
\begin{eqnarray}
\Vert [a_y, \alpha^{\Lambda}_t(B)]\Vert
\leq 
\Vert a_y\Vert
\sum_{x \in {\bf Z}} \sup_{c \in {\cal A}(\{0\})}
\left( \frac{\Vert [\tau_{x+y}(c),B]\Vert}
{\Vert c\Vert} \right) e^{-|x|\lambda +2|t|\Vert \Phi \Vert_{\lambda}}
\end{eqnarray}
holds.
By letting $\Lambda$ to ${\bf Z}$, we obtain 
\begin{eqnarray}
\frac{\Vert [a_y, \alpha_t(B)]\Vert}{\Vert a_y \Vert}
\leq 
\sum_{x \in {\bf Z}} F_{x+y}(B)
e^{-|x|\lambda +2|t|\Vert \Phi \Vert_{\lambda}}.
\end{eqnarray}
Thus 
\begin{eqnarray}
F_y(\alpha_t(B)) 
&\leq& 
\sum_{x \in {\bf Z}} F_{x+y}(B)
e^{-|x|\lambda +2|t|\Vert \Phi \Vert_{\lambda}}
\nonumber \\
&\leq &
\sum_{z  \in {\bf Z}}F_{z}(B)
e^{-|z-y|\lambda +2|t|\Vert \Phi \Vert_{\lambda}}
\nonumber \\
&=&
\sum_{z\geq y}F_z(B) e^{-(z-y)\lambda +2|t|\Vert \Phi\Vert_{\lambda}}
+
\sum_{z < y}F_z(B) e^{-(y-z)\lambda +2|t|\Vert \Phi\Vert_{\lambda}}
\end{eqnarray}
holds. Since we are interested in asymptotic behavior for 
large $|y|$, let us consider the case when $y$ is positively large.
For any $\epsilon_0 >0$, there exists $N>0$ such that for 
$|z|>N$, 
$F_z(B)<\epsilon_0 e^{-\mu |z|}$ holds. By use of 
these relations we decompose the above equation.
Let us choose $y$ which is larger than $N$, then 
the first term is bounded as
\begin{eqnarray}
\sum_{z\geq y}F_z(B) e^{-(z-y)\lambda +2|t|\Vert \Phi\Vert_{\lambda}}
&<& \epsilon_0 \sum_{z\geq y}e^{-\mu z}
 e^{-(z-y)\lambda +2|t|\Vert \Phi\Vert_{\lambda}}
\nonumber \\
&=&\epsilon_0 \frac{e^{-\mu y}}{1-e^{-(\mu+\lambda)}}
e^{2|t|\Vert \Phi\Vert_{\lambda}}.
\end{eqnarray}
The second term is also decomposed and bounded as 
\begin{eqnarray}
\sum_{z < y}F_z(B) e^{-(y-z)\lambda +2|t|\Vert \Phi\Vert_{\lambda}}
&=&\sum_{z \in (-N,N)}
F_z(B) e^{-(y-z)\lambda +2|t|\Vert \Phi\Vert_{\lambda}}
+
\sum_{z \leq -N}F_z(B) e^{-(y-z)\lambda +2|t|\Vert \Phi\Vert_{\lambda}}
\nonumber \\
&+&
\sum_{N\leq z < y}F_z(B) e^{-(y-z)\lambda +2|t|\Vert \Phi\Vert_{\lambda}}
\nonumber \\
&\leq& 
e^{-\lambda y}\sum_{z \in (-N,N)}
F_z(B) e^{z\lambda +2|t|\Vert \Phi\Vert_{\lambda}}
+
\epsilon_0 
e^{-\lambda y}\frac{e^{-(\lambda +\mu)N}}{1-e^{-(\lambda+\mu)}}e^{2|t|
\Vert \Phi\Vert_{\lambda}}
\nonumber \\
&+&
\epsilon_0 e^{-\lambda y}
\frac{e^{N(\lambda-\mu)}}
{1-e^{\lambda-\mu}}e^{2|t|\Vert \Phi\Vert_{\lambda}}
-
\epsilon_0 e^{-\mu y}
\frac{1}
{1-e^{\lambda-\mu}}e^{2|t|\Vert \Phi\Vert_{\lambda}}.
\end{eqnarray}
Thus for sufficiently small $\epsilon'$, 
\begin{eqnarray}
\lim_{y \to \infty} e^{\epsilon' y} F_y(\alpha_t(B)) =0
\end{eqnarray}
holds. With respect to the asymptotic behavior for $y \to -\infty$,
the proof goes similarly.
\hfill Q.E.D.
\par 
We, hereafter, consider the dynamical entropy 
$h(\omega,\alpha,{\cal A}_{\exp})$
for a translationally invariant stationary state $\omega$.
\section{Bound for Dynamical Entropy of Spin Systems}\label{sect:bound}
In this section we bound the dynamical entropy of quantum spin systems.
It seems natural to imagine that the range of the interaction has relationship 
with the dynamical entropy. For instance, if the potential $\Phi(X)$ 
vanishes for $|X|\geq 2$ and the partition of unity is strictly local,
the state can be disturbed only for the fixed finite region.
Thus the dynamical entropy is vanishing in such a case. 
Such an observation guides us to a conjecture that 
in general case the quantum dynamical entropy is related with 
the range of the interaction through its corresponding 
group velocity. 
\par
First let us begin with a lemma approximating an element 
$A \in {\cal A}_{\exp}$ by strictly local objects.
As was introduced in \cite{FT}, we introduce a conditional expectation
on ${\cal A}([-L,L])\ (L>0)$ by
\begin{eqnarray}
id_{[-L,L]}\otimes \tau_{[-L,L]^c}
\end{eqnarray}
where $\tau_{[-L,L]^c}$ 
is normalized trace on $[-L,L]^c={\bf Z}\setminus [-L,L]$.
For $A \in {\cal A}_{\exp}$, an estimate,
\begin{eqnarray} 
\Vert A-id_{[-L,L]}\otimes \tau_{[-L,L]^c}(A)\Vert \leq \sum_{
x \in [-L,L]^c} F_x(A)
\end{eqnarray}
holds \cite{FT}.
Now for any $\epsilon_0$, there exists $M>0$ such that 
for all $|x|>M$, $F_{x}(A) <\epsilon_0 e^{-\mu |x|}$
holds.
If we take $L$ as $L>M$ we obtain
\begin{eqnarray}
\sum_{
x \in [-L,L]^c} F_x(A)
< \frac{2 \epsilon_0}{1-e^{-\mu}}e^{-\mu L}.
\end{eqnarray}
Thus the following lemma holds.
\begin{lemma}\label{lemma1}
For any $A \in {\cal A}_{\exp}$
and $\epsilon_1>0$, there exists $M>0$ such that 
the following condition is satisfied.
For any $L>M$, there exists a strictly local object $A_{L} \in 
{\cal A}([-L,L])$ such that $\Vert A-A_L \Vert < \epsilon_1 e^{-\mu L}$
and $\Vert A_L\Vert \leq \Vert A\Vert$ holds.
\end{lemma}
Thus we obtained a good approximation of 
exponentially localized object by local ones.
Although the strictly local object does not remain
strictly local as time elapses, we can prove the following lemma.
\begin{lemma}\label{lemma2}
For any strictly local object $A\in {\cal A}([-L,L])$, $t \in {\bf R}$
 and $\epsilon_2 >0$, if $R \in {\bf Z}$ satisfies
\begin{eqnarray}
R> L+
\frac{2 \Vert \Phi \Vert_{\lambda}}{\lambda}|t|
+\frac{1}{\lambda}\left\{
\log\left( \frac{2e^{-\lambda}}{1-e^{-\lambda}}\right)
+4(2L+1)\log(N+1) +\log (2L+1) -\log \epsilon_2\right\},
\end{eqnarray}
an inequality
\begin{eqnarray}
\Vert \alpha_t (A)-\alpha^{[-R,R]}_t(A)\Vert 
\leq \epsilon_2 \Vert A\Vert
\end{eqnarray}
holds.
\end{lemma}
{\bf Proof:}
$A \in {\cal A}([-L,L])$ can be decomposed into 
\begin{eqnarray}
A=\sum_{\{i_x\},\{j_x\}}
C(\{i_x\},\{j_x\})
\Pi_{x \in [-L,L]}
e(i_x,j_x)
\end{eqnarray}
where $\{e(i_x,j_x)\}\subset {\cal A}(\{x\}),(i_x,j_x
=0,1,\cdots,N)$ is a set of matrix elements.
Then we must bound 
\begin{eqnarray}
\Vert \alpha_t(A)-\alpha_t^{[-R,R]}(A)\Vert
&\leq &
 \sum_{\{i_x\},\{j_x\}}
 |C(\{i_x,j_x\})|\Vert 
 \Pi_{x \in [-L,L]}
\alpha_t(e(i_x,j_x))
-
 \Pi_{x \in [-L,L]}
 \alpha_t^{[-R,R]}(e(i_x,j_x))\Vert
\nonumber \\
&\leq &
 \sum_{\{i_x\},\{j_x\}}
 |C(\{i_x,j_x\})|
 \sum_{x \in [-L,L]}
 \Vert 
 \alpha_t(e(i_x,j_x))
-\alpha_t^{[-R,R]}(e(i_x,j_x))
\Vert.
\nonumber
\end{eqnarray}
Due to theorem 6.2.11 in \cite{BR},
for $x \in [-L,L]$,
we can apply the following inequality, 
\begin{eqnarray}
 \Vert 
 \alpha_t(e(i_x,j_x))
-\alpha_t^{[-R,R]}(e(i_x,j_x))
\Vert
\leq (2L+1)\frac{2e^{-\lambda}}{1-e^{-\lambda}}
e^{-\lambda (R-L) +2 |t|\Vert \Phi \Vert_{\lambda}}.
\end{eqnarray}
We obtain,
\begin{eqnarray}
&&
\sum_{\{i_x\},\{j_x\}}
 |C(\{i_x,j_x\})|
 \sum_{x \in [-L,L]}
 \Vert 
 \alpha_t(e(i_x,j_x))
-\alpha_t^{[-R,R]}(e(i_x,j_x))
\Vert
\nonumber \\
&&\leq 
\sum_{\{i_x\},\{j_x\}}
 |C(\{i_x,j_x\})|
 (2L+1)\frac{2e^{-\lambda}}{1-e^{-\lambda}}
e^{-\lambda (R-L) +2 |t|\Vert \Phi \Vert_{\lambda}}
\nonumber \\
&&\leq
\Vert A\Vert (N+1)^{2(2L+1)}
(2L+1)\frac{2e^{-\lambda}}{1-e^{-\lambda}}
e^{-\lambda (R-L) +2 |t|\Vert \Phi \Vert_{\lambda}},
\end{eqnarray}
where we used $|C(\{i_x,j_x\})|\leq \Vert A\Vert$.
\hfill Q.E.D.
\par
Now we can prove the main theorem.
\begin{theorem}
For spin systems, the dynamical entropy with respect to 
a translationally invariant stationary state $\omega$, 
is bounded from above by 
the following inequality,
\begin{eqnarray}
h(\omega,\alpha,{\cal A}_{\exp})\leq 2V(\Phi) \left( \sigma(\omega)+\log(N+1) 
\right),
\end{eqnarray}
where $\sigma(\omega):=\lim_{\Lambda \to {\bf Z}}
\frac{S(\omega|_{\Lambda})}{|\Lambda|}$ is a mean entropy and 
$V(\Phi):=\inf_{\lambda}\left( \frac{\Vert \Phi\Vert_{\lambda}}
{\lambda}\right)$ is a quantity called group velocity.
\end{theorem}
{\bf Proof:} 
Let us consider the dynamical entropy for a
fixed partition of unity $\{x_i\}_{i=1}^Z \subset {\cal A}_{\exp}$.
The strategy is to approximate $\alpha_t(x_i)$ by 
strictly local object. For any $\epsilon_1 >0$ and
$\epsilon_2>0$, there exists $M>0$ such that 
for $L\geq M$ one can choose $x_i^L\in {\cal A}([-L,L])$ 
with
$\Vert x_i^L\Vert \leq \Vert x_i\Vert$
satisfying
\begin{eqnarray}
\Vert \alpha_t(x_i)-\alpha^{[-R,R]}_t(x^{L}_i) \Vert
&\leq&
\Vert x_i - x^{L}_i \Vert + \Vert
\alpha^{[-R,R]}_t(x_i^L) -\alpha_t (x_i^L)\Vert
\nonumber \\
&\leq&
\epsilon_1 e^{-\mu L} +\epsilon_2 \Vert x_i\Vert,
\end{eqnarray}
where $R \in {\bf Z}$ is fixed by
\begin{eqnarray}
R&\geq& L+
\frac{2 \Vert \Phi \Vert_{\lambda}}{\lambda}|t|
+\frac{1}{\lambda}\left\{
\log\left( \frac{2e^{-\lambda}}{1-e^{-\lambda}}\right)
+4(2L+1)\log(N+1) +\log (2L+1) -\log \epsilon_2\right\} 
\nonumber\\
&>&R-1.
\label{abovecond}
\end{eqnarray}
For any $\epsilon>0$, we fix $\epsilon_1$ and $\epsilon_2$
by
\begin{eqnarray}
\epsilon_1 &=&\frac{\epsilon}{2}
\nonumber \\
\epsilon_2 &=&
\frac{\epsilon}{2 t^2 \Vert x_i \Vert^2}.
\label{Ri}
\end{eqnarray}
Then there exists $M>0$ such that for $L\geq M$
one can find $x_i^L$ satisfying 
\begin{eqnarray}
\Vert \alpha_t(x_i)-\alpha^{[-R_i,R_i]}_t(x^{L}_i) \Vert
&\leq& 
\frac{\epsilon}{2}(e^{-\mu L}+\frac{1}{t^2}),
\end{eqnarray}
where $R=R_i$ is determined by the above condition (\ref{abovecond}) with 
(\ref{Ri}).
Thus if we fix $L$ by
\begin{eqnarray}
L\geq \max\{\frac{2\log t}{\mu},M\} \geq L-1, 
\end{eqnarray}
we obtain a bound,
\begin{eqnarray}
\Vert \alpha_t(x_i)-\alpha^{[-R_i,R_i]}_t(x^{L}_i) \Vert
< \frac{\epsilon}{t^2}. \label{app}
\end{eqnarray}
As a result, for sufficiently large $t>0$ we can 
approximate $\alpha_t(x_i)$ by a strictly local 
element $\alpha_t^{[-R,R]}(x_i^L)\in {\cal A}([-R,R])$
with 
\begin{eqnarray}
\Vert \alpha_t(x_i)-\alpha^{[-R,R]}_t(x^{L}_i) \Vert
< \frac{\epsilon}{t^2},
\end{eqnarray}
where $R$ is determined by $R:=\max\{R_i\}$ 
and
\begin{eqnarray}
R_i\geq && \frac{2\log t}{\mu}+
\frac{2 \Vert \Phi \Vert_{\lambda}}{\lambda}|t|
+
\frac{1}{\lambda}\left\{
\log\left( \frac{2e^{-\lambda}}{1-e^{-\lambda}}\right)
+4(2\frac{2\log t}{\mu}+1)\log(N+1)\right.
\nonumber \\
&& 
\left.
+\log (2\frac{2\log t}{\mu}
+1) -\log \epsilon +\log 2 
+ 2 \log |t| +2 \log \Vert x_i \Vert\right\}
\nonumber\\
>&&R_i-1.
\end{eqnarray}
Thanks to the lemma 3.3 in \cite{FT},
for sufficiently small $\epsilon>0$,
we obtain
\begin{eqnarray}
\left| 
S(\rho(\{x_i\}_i,t,\alpha_1))
-S(\rho(\{x^L_i\},t;\alpha^{[-R,R]}_1)
\right|
\leq
2Z \log (2Z^t)\frac{\epsilon \pi^2}{6}
-2Z \left(\frac{\epsilon \pi^2}{6} \right)
\log \left( 2Z \frac{\epsilon \pi^2}{6}\right).
\label{1}
\end{eqnarray}
Now thanks to the lemma 3.2 in \cite{FT},
$S(\rho(\{x^L_i\},t;\alpha^{[-R,R]}_1)$
can be bounded as
\begin{eqnarray}
S(\rho(\{x^L_i\},t;\alpha^{[-R,R]}_1)
\leq S(\left. \omega\right|_{[-R,R]})
+(2R+1) \log (N+1),
\label{2}
\end{eqnarray}
where $\left. \omega\right|_{[-R,R]}$ 
is a restriction of the state $\omega$ on 
${\cal A}([-R,R])$ and $S(\left. \omega\right|_{[-R,R]})$ 
is its von Neumann entropy.
Combination of (\ref{1}) and (\ref{2}) and taking a limit 
of 
$\epsilon \to 0$ we finally obtain
\begin{eqnarray}
h(\omega,\alpha,{\cal A}_{\exp})
\leq 2 \frac{2\Vert \Phi\Vert_{\lambda}}{\lambda}
\left(\sigma(\omega)+\log(N+1)
\right),
\end{eqnarray}
where $\sigma(\omega):=\lim_{\Lambda \to {\bf Z}}\frac{1}{|\Lambda|}
S(\left. \omega\right|_{\Lambda})$ is a mean entropy.
\hfill Q.E.D.
\section{Conclusion and Outlook}
In this paper we investigated the quantum dynamical entropy 
of one-dimensional quantum spin systems. A generalization of 
our result to higher dimensional lattices is 
straightforward. The upper bound we found 
seems to be natural since the quantum dynamical entropy roughly represents
the rate of how large subalgebra of the system 
can be concerned as time elapses, and
the group velocity bounds the region one can disturb.  
Our result does not depend upon the details of 
the interaction, and is rather general one.
One might wonder whether it is possible to obtain 
a lower bound for the spin systems. It will depend upon 
the form of the interaction since the ergodic property 
will be strongly related. Estimations of the other 
dynamical entropies on the spin systems are interesting.
We will address  
these problems elsewhere.
\\
{\bf Acknowledgments}
\\
T.M would like to thank Prof. Robert Alicki for his remarks and 
encouragement.


\begin{thebibliography}{99}

\bibitem{CS}A.Connes and E.Stormer, Entropy for automorphisms of 
$\mbox{II}_1$ von Neumann algebras,
Acta Math. {\bf 134} (1975) 289-306.

\bibitem{CNT}A.Connes, H.Narnhofer and W.Thirring,
Dynamical entropy of $\mbox{C}^*$-algebras and von Neumann algebras,
Commun.Math.Phys.{\bf 112} (1987) 691-719.

\bibitem{AF}R.Alicki and M.Fannes, Defining quantum dynamical entropy,
Lett.Math.Phys.{\bf 32} (1994) 75-82.

\bibitem{AFbook}R.Alicki and M.Fannes,
{\it Quantum Dynamical Systems}, Oxford Univ. Press
(2001)

\bibitem{MO}N.Muraki and M.Ohya,
Entropy functionals of Kolmogorov-Sinai type and their
limit theorems,
Lett.Math.Phys. {\bf 36} (1996) 327-335.

\bibitem{AOW}L.Accardi, M.Ohya and N.Watanabe,
Dynamical entropy through quantum Markov chains,
Open System Infor. Dynamics {\bf 4} (1997),71-87.

\bibitem{KOW}A.Kossakowski, M.Ohya and N.Watanabe,
Quantum dynamical entropy for completely positive maps,
Infinite Dim. Analysis, Quantum Prob. Related Topics, {\bf 2}
2 (1999) 267-282.

\bibitem{AOW2}L.Accardi, M.Ohya and N.Watanabe,
Note on quantum dynamical entropies,
Rep.Math.Phys.{\bf 38} (1996), 457-469.

\bibitem{AN}R.Alicki and H.Narnhofer,
Comparison of dynamical entropies of non-commutative shift,
Lett.Math.Phys., {\bf 33}, (1995) 241-247.

\bibitem{AFTA}J.Andries, M.Fannes, P.Tuyls and R.Alicki,
The dynamical entropy of the quantum Arnold cat map,
Lett.Math.Phys. {\bf 35}, (1995) 375-383

\bibitem{FT}M.Fannes and P.Tuyls,
A continuity property of quantum dynamical entropy,
Infinite Dim. Analysis, Quantum Prob. Related Topics, {\bf 2}
(1999) 511-527.

\bibitem{OP}M.Ohya and D.Petz,
{\it Quantum Entropy and Its Use},Texts and Monographs in Physics.
Springer-Verlag (1993)

\bibitem{AFL}L.Accardi, A.Frigerio and J.Lewis,
Quantum stochastic processes,
Publ.RIMS. {\bf 18} (1982) 97.

\bibitem{FNW}M.Fannes, B.Nachtergaele and R.F.Werner,
Finitely correlated states on quantum spin chains,
Commun.Math.Phys. {\bf 144}, (1992) 443-490

\bibitem{BR}O.Bratteli and D.W.Robinson, {\it %
Operator algebras and quantum statistical mechanics. 2. Equilibrium states.
Models in quantum statistical mechanics.} Texts and Monographs in Physics.
(1981)
Springer-Verlag, Berlin.

\end{thebibliography}
\end{document}